\documentclass{aa}
\usepackage{subcaption}

\begin{document}
\title{Resonant absorption in expanding coronal magnetic flux tubes with uniform density}
\author{T. A. Howson \inst{1} \and I. De Moortel \inst{1,2} \and P. Antolin \inst{1} \and T. Van Doorsselaere \inst{3} \and A. N. Wright \inst{1}}
\institute{School of Mathematics and Statistics, University of St. Andrews, St. Andrews, Fife, KY16 9SS, U.K. \and Rosseland Centre for Solar Physics, University of Oslo, PO Box 1029  Blindern, NO-0315 Oslo, Norway \and Centre for mathematical Plasma Astrophysics, Department of Mathematics, KU Leuven, Celestijnenlaan 200B, 3001 Leuven, Belgium}

\abstract{}
{We investigate the transfer of energy between a fundamental standing kink mode and azimuthal Alfv\'en waves within an expanding coronal magnetic flux tube. We consider the process of resonant absorption in a loop with a non-uniform Alfv\'en frequency profile but in the absence of a radial density gradient.}
{Using the three dimensional MHD code, Lare3d, we modelled a transversely oscillating magnetic flux tube that expands radially with height. An initially straight loop structure with a magnetic field enhancement was allowed to relax numerically towards a force-free state before a standing kink mode was introduced. The subsequent dynamics, rate of wave damping and formation of small length scales are considered.}  
{We demonstrate that the transverse gradient in Alfv\'en frequency required for the existence of resonant field lines can be associated with the expansion of a high field-strength flux tube from concentrated flux patches in the lower solar atmosphere. This allows for the conversion of energy between wave modes even in the absence of the transverse density profile typically assumed in wave heating models. As with standing modes in straight flux tubes, small scales are dominated by the vorticity at the loop apex and by currents close to the loop foot points. The azimuthal Alfv\'en wave exhibits the structure of the expanded flux tube and is therefore associated with smaller length scales close to the foot points of the flux tube than at the loop apex.}
{Resonant absorption can proceed throughout the coronal volume, even in the absence of visible, dense, loop structures. The flux tube and MHD waves considered are difficult to observe and our model highlights how estimating hidden wave power within the Sun's atmosphere can be problematic. We highlight that, for standing modes, the global properties of field lines are important for resonant absorption and coronal conditions at a single altitude will not fully determine the nature of MHD resonances. In addition, we provide a new model in partial response to the criticism that wave heating models cannot self-consistently generate or sustain the density profile upon which they typically rely.}
{}
\maketitle


\section{Introduction}\label{sec:introduction}
In the years since the launch of the TRACE (Transition Region and Coronal Explorer) mission in 1998, high spatial and temporal resolution imaging has enabled the detection of a multitude of oscillations within the solar atmosphere \citep[e.g.][]{Aschwanden1999,  Okamoto2007, Banerjee2009, Marsh2011, Freij2014}. Indeed, more recent studies using, for example, the CoMP (Coronal Multi-channel Polarimeter) coronagraph have highlighted the apparent abundance of wave power throughout the coronal volume \citep{Tomczyk2007, Morton2016, Morton2019}. Whilst estimates of the energy associated with these waves are not typically well constrained (in part due to the difficulty of detecting incompressible and small scale wave modes), some authors suggest that it is sufficient to heat the (quiescent) corona \citep{McIntosh2011, Morton2012} and accelerate the fast solar wind \citep{DePontieu2007}.  

Despite the expected rate of wave energy dissipation being very low within the corona, many observations of transverse loop oscillations exhibit rapid damping, over the course of a few wave periods \citep[e.g.][]{Nakariakov1999, Schrijver2000}. This is now widely understood to be caused by the well-studied process of resonant absorption \citep{Ionson1978, Goossens2002}, by which energy is transferred from the easily observed kink mode to localised, azimuthal, Alfv\'en waves that are much more difficult to detect \citep{Pascoe2010, DeMoortel2012}. Since this is an ideal process, no wave energy is dissipated during the conversion of wave modes. However, the resultant Alfv\'en waves are associated with much smaller length scales and may experience enhanced dissipation as a result of phase mixing \citep{Heyvaerts1983, Parker1991, Pagano2017}.

Since these Alfv\'en modes are only weakly compressible (fully incompressible for the $m=0$ mode), they remain difficult to detect within the solar corona. However, in recent years, several authors have interpreted torsional oscillatory motions as evidence for the presence of Alfv\'en waves in the solar atmosphere \citep[e.g.][]{Tomczyk2007, Jess2009, Srivastava2017}. Furthermore, synthetic observations derived by forward modelling the results of numerical simulations, have highlighted the signatures of resonant absorption \citep{Goossens2014, Karampelas2019}, phase mixing and the associated growth of the Kelvin-Helmholtz instability \citep{Terradas2008a, Antolin2014} that could be detected given current instrumental constraints \citep{Antolin2017}.

Within an oscillating coronal loop, the transfer of energy from the kink mode to the azimuthal Alfv\'en mode relies on the existence of a set of magnetic field lines with natural Alfv\'en frequencies that coincide with the frequency of the kink wave \citep[e.g.][]{Sakurai1991}. The kink speed for a thin magnetic flux tube is given by a density-weighted mean of the internal and external Alfv\'en speeds \citep[see, for example,][]{Priest2014}. Hence, if the Alfv\'en speed varies smoothly across the loop, a resonance will occur somewhere within the loop's radius. As mentioned above, the enhanced energy dissipation rate associated with the excited wave modes are of interest in the context of coronal heating. However, \citet{Cargill2016} argued that even if significant wave heating occurs, energy will only be dissipated in narrow regions in the boundary of a coronal loop and thus, will not be able to sustain (or create) the typically-assumed density profile. Despite this, recent numerical studies have highlighted the possibility that non-linear effects can cause heating throughout the cross-section of a dense flux tube \citep[e.g.][]{Karampelas2018}. 

Typically, in coronal wave models, the radial variation in the Alfv\'en speed is associated with a non-uniform transverse density profile. However, a radial gradient in the magnetic field strength can also provide the non-constant Alfv\'en speed profile that is required for a resonant layer to exist. To this end, many studies have demonstrated the mode conversion of MHD waves in the presence of a non-uniform magnetic field \citep[see, for example,][]{Ruderman2009, Antolin2015, Giagkiozis2016, Yu2017B, Howson2017B}. However, in the majority of previous studies, the magnetic field is simply a function of radial position and does not vary along the length of the magnetic structure. In the current study, the magnetic flux tube expands with height in the corona and, thus, the background field strength is no longer constant along field lines. Importantly, the ratio between the internal and external Alfv\'en speeds (and hence the kink speed) is not constant along the length of the loop.

In the context of magnetospheric waves, previous studies have demonstrated the existence of resonances associated with varying magnetic field strength. An analytic treatment of resonances forming in fields with an invariant direction can be found in \citet{Wright1994} and more recent studies \citep{Wright2016, Elsden2017} have explored the existence of resonant regions in fully three dimensional (no invariant direction) magnetic fields. 

Within the Earth's approximately dipolar magnetosphere, the field strength falls with height, modifying the natural frequency of magnetic field lines from those in a uniform field regime. An analogy may be drawn with the solar atmosphere in which flux tubes might be expected to expand rapidly with height in the transition region and corona from dense flux patches in the photosphere and chromosphere. A similar regime is explored in \citet{Khomenko2008} in which the authors explore the dynamics of propagating magnetoacoustic waves in small, expanding photospheric flux tubes. 

In other related studies, many authors have demonstrated that resonant absorption will occur in a wide array of coronal loop-like structures. In particular, the process will still progress in multi-stranded \citep{Terradas2008b}, curved \citep{VanDoors2004, Terradas2006},  elliptical \citep{Ruderman2003} and longitudinally stratified loops \citep{Andries2005, Arregui2005}.

Within this paper, we aim to establish the possibility of resonant absorption occuring in magnetic loop-like structures that are bereft of any density enhancement. We introduce the model in Sect. 2, describe our results in Sect. 3 and provide a discussion and conclusions in Sect. 4.


\section{Numerical Method}
For the numerical simulations presented within this paper, we have used the Lagrangian-remap code, Lare3D \citep{Arber2001}. We advanced the full, 3-D, ideal MHD equations in normalised form given by

\begin{equation}\frac{\text{D}\rho}{\text{D}t} = -\rho \vec{\nabla} \cdot \vec{v}, \end{equation}
\begin{equation} \label{eq:motion} \rho \frac{{\text{D}\vec{v}}}{{\text{D}t}} = \vec{j} \times \vec{B} - \vec{\nabla} P, \end{equation}
\begin{equation} \label{eq:energy} \rho \frac{{\text{D}\epsilon}}{{\text{D}t}} = - P(\vec{\nabla} \cdot \vec{v}), \end{equation}
\begin{equation}\frac{\text{D}\vec{B}}{\text{D}t}=\left(\vec{B} \cdot \vec{\nabla}\right)\vec{v} - \left(\vec{\nabla} \cdot \vec{v} \right) \vec{B}. \end{equation}

Here, $\rho$ is the plasma density, $\vec{v}$ is the velocity field, $\vec{j}$ is the current density, $\vec{B}$ is the magnetic field, $P$ is the gas pressure and $\epsilon$ is the specific internal energy density. This normalisation is constructed using a typical length scale, magnetic field strength and density and details of the code are described in \citep{Arber2001}. The results presented hereafter are calculated using S.I. (not dimensionless) units unless otherwise stated. Although we display the ideal MHD equations here, we note that there is a small viscosity included in the code to ensure numerical stability. This is included as a small, dissipative force on the right-hand side of the equation of motion (\ref{eq:motion}) and an associated heating term is added to the right-hand side of the energy equation (\ref{eq:energy}). Although the dissipative effects are weak here, even small transport coefficients can prevent the growth of the Kelvin-Helmholtz instability that may develop as a result of the radial velocity shear that forms during the simulation \citep{Howson2017a}. 

We consider an azimuthally-invariant cylindrical structure with uniform density and an internal enhancement of the magnetic field. This will be associated with a radial magnetic pressure force and in order to find an initial equilibrium, we consider solutions to equation \ref{eq:motion}, with $\vec{v}=\vec{0}$. Hence,

\begin{equation} \label{eq:balance} \vec{\nabla}P = \vec{j} \times \vec{B} = \underbrace{\left(\vec{B} \cdot \vec{\nabla}\right)\vphantom{\left(\frac{B^2}{2}\right)}\vec{B}}_{\text{\shortstack{Magnetic \\ 
Tension}}} - \underbrace{\vec{\nabla}\left(\frac{B^2}{2}\right).}_{\text{\shortstack{Magnetic\\ 
Pressure}}}\end{equation}

We explore three cases for satisfying equation \ref{eq:balance}.

\begin{enumerate} 
\item {\emph{External gas pressure enhancement}} - Requires $\beta > 1$ in the external medium which is not suitable for a coronal investigation. However, this may represent a flux tube located within the lower layers of the solar atmosphere \citep[see, for example,][]{Yu2017B}.

\item {\emph{Twisted field}} - The radially outwards magnetic pressure force can be balanced by an inwards magnetic tension force associated with twisted field. The presence of twist within the magnetic field can complicate the wave dynamics \citep[see e.g.][]{Karami2009, Howson2017B} and so is not suitable for an initial study.

\item {\emph{Field expansion}} - An initially straight flux tube can be allowed to relax numerically to a state with $\vec{j} \times \vec{B} = \vec{0}$. The magnetic field strength will inevitably decrease within the flux tube as it expands, however, it can be constrained to a limited degree by tension in the field lines. 
\end{enumerate}

For the remainder of this article, we will restrict our consideration to the third case.
\begin{figure}[h]
  \centering
  \includegraphics[width=0.5\textwidth]{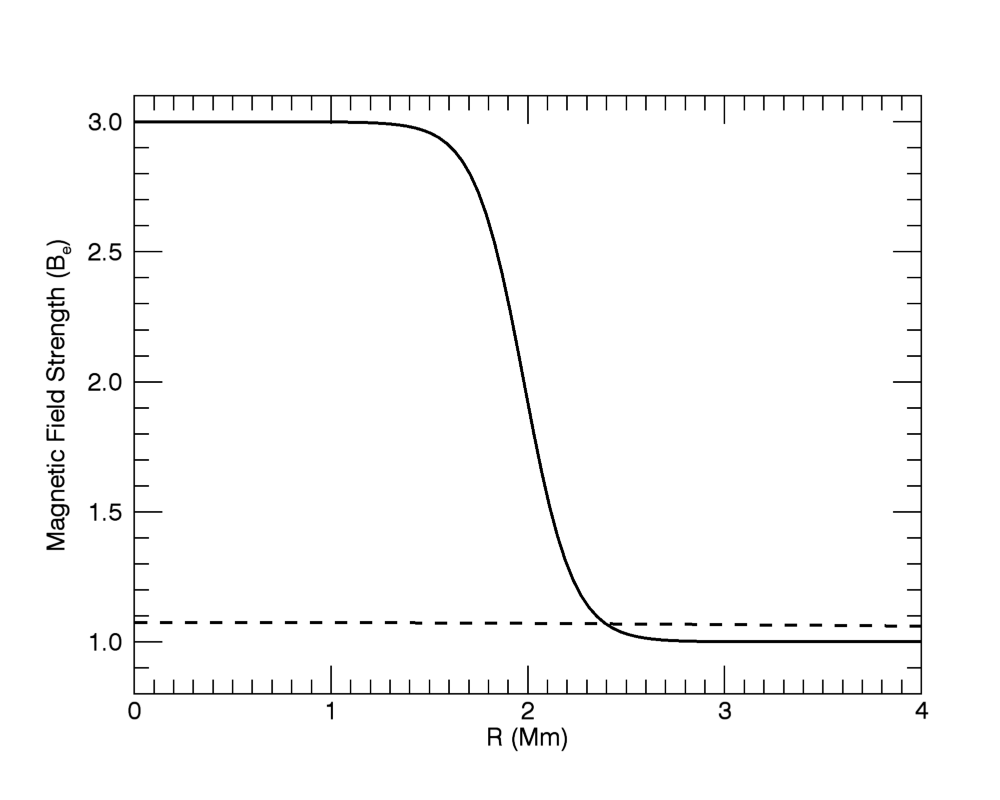}
  \caption{Initial magnetic field strength profile through the cross-section of the loop. The solid line shows the pre-relaxation field profile and the post-relaxation field profile at both of the $z$ boundaries. The dashed line shows the field profile at the loop apex following the relaxation. In both cases we have normalised by the initial external field strength and note that the entire radial extent of the domain is not included within this figure.}
  \label{InField}
\end{figure}

\subsection{Initial Set-up}
The initial conditions consisted of a straight, vertically and azimuthally invariant magnetic flux tube with a field parallel to the loop axis. We imposed an enhanced field strength within the loop (relative to its exterior) of the form $\vec{B} = (0,0, B_z)$ where $z$ is the loop-aligned coordinate and

\begin{equation}
B_z(R) = B_e + \frac{\left(B_i -B_e\right)}{2}\left(1-\tanh\left\{\frac{R-r_a}{r_b}\right\}\right).
\end{equation}

Here, $B_e = 5$ G and $B_i = 15$ G are the initial exterior and interior field strengths, respectively. The parameters, $r_a$ and $r_b$ were set to produce a loop radius of approximately 2 Mm and a smooth transition from the exterior field to the interior field  of approximately 0.8 Mm in width. The radial profile of this field is shown (solid line) in Fig. \ref{InField}. The loop length was 20 Mm. We note that this is relatively short for a coronal flux tube and discuss the significance of this length below (see Sect. \ref{Discussion}).

We used a numerical domain of $512 \times 512 \times 200$ grid cells and in order to minimise boundary effects, we adopted a non-uniform resolution profile in the $x$ and $y$ directions. These profiles have a region of uniform resolution in the centre of the domain in which the important wave dynamics (e.g. resonant absorption) occur \citep[see][for more details]{Howson2017a}. The use of the non-uniform grid allowed a fine spatial resolution  (0.04 Mm) to be obtained within a central region whilst ensuring the $x$ and $y$ edges are both 32 Mm in length. The large distance of the $x$ and $y$ boundaries from the centre of the flux tube ensured that during relaxation (see below), the magnetic field was not artificially constrained by the boundaries of the domain. Meanwhile, the $z$ direction used a uniform spatial resolution of 100 km. Unlike in simulations of straight flux tubes, the expansion of the magnetic structure discussed within this article (see below) presents some numerical difficulties. In particular, a significantly finer spatial resolution is required close to the loop foot points in order to resolve the gradients associated with the rapid expansion of the flux tube. 

Whilst maintaining a density of $1.67 \times 10^{-12} \text{ kg m}^{-3}$ and a temperature of 1.8 MK throughout the computational domain, we allowed the magnetic field to relax towards a numerical equilibrium. A high value of viscosity was implemented to damp the amplitude of the oscillations that form, however, the associated heating was removed by overwriting the temperature at each time step. During the numerical relaxation, we maintained the initial magnetic field profile on the upper and lower $z$ boundaries. Following this process, the magnetic field (numerically) satisfied $\vec{j} \times \vec{B} =\vec{0}$ with the exception of a narrow layer close to the top and bottom of the domain where the boundary conditions were associated with non-parallel currents. 

\begin{figure}[h]
  \centering
  \includegraphics[width=0.4\textwidth]{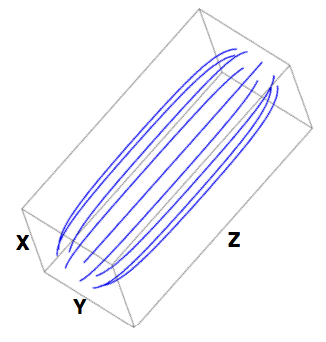}
  \caption{Magnetic field lines traced from $R=2$ Mm on the lower $z$ boundary after the numerical relaxation. The displayed box does not show the entirety of the computational domain but, instead, has dimensions of 8 Mm $\times$ 8 Mm $\times$ 20 Mm.}
  \label{Field_Profiles}
\end{figure}

\begin{figure}[h]
  \centering
  \includegraphics[width=0.5\textwidth]{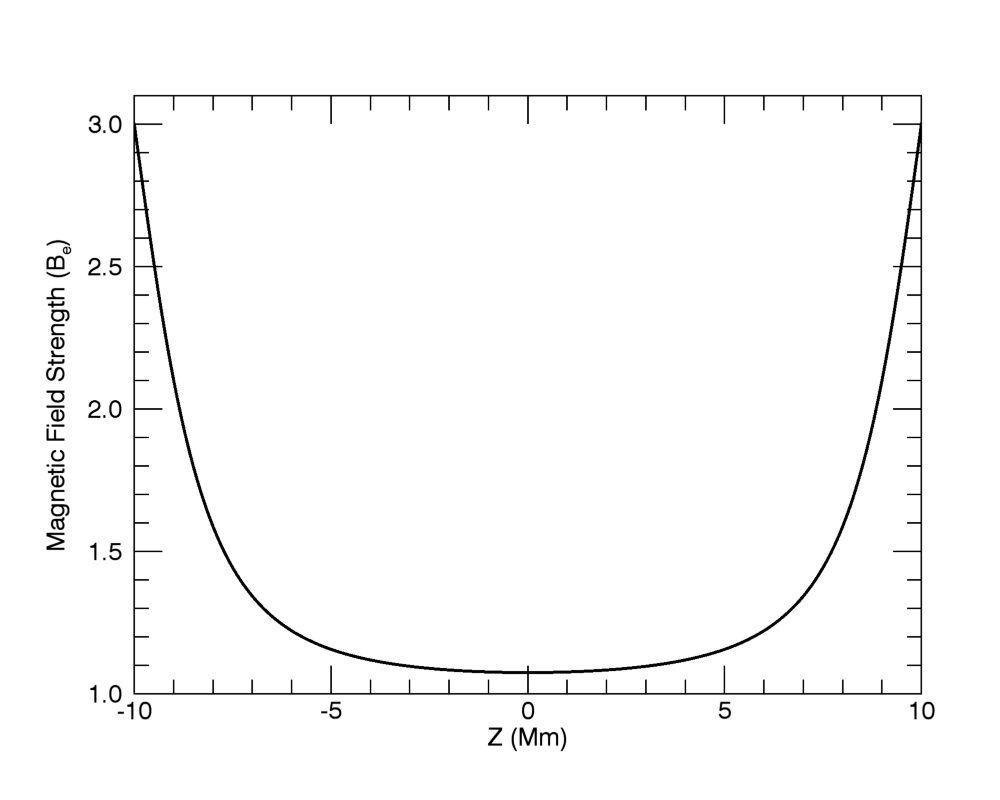}
  \caption{Field strength variation along the flux tube axis (central field line) after the period of numerical relaxation. Here, we have normalised by the initial external field strength.}
  \label{InField_Vertical}
\end{figure}

\begin{figure}[h]
  \centering
  \includegraphics[width=0.5\textwidth]{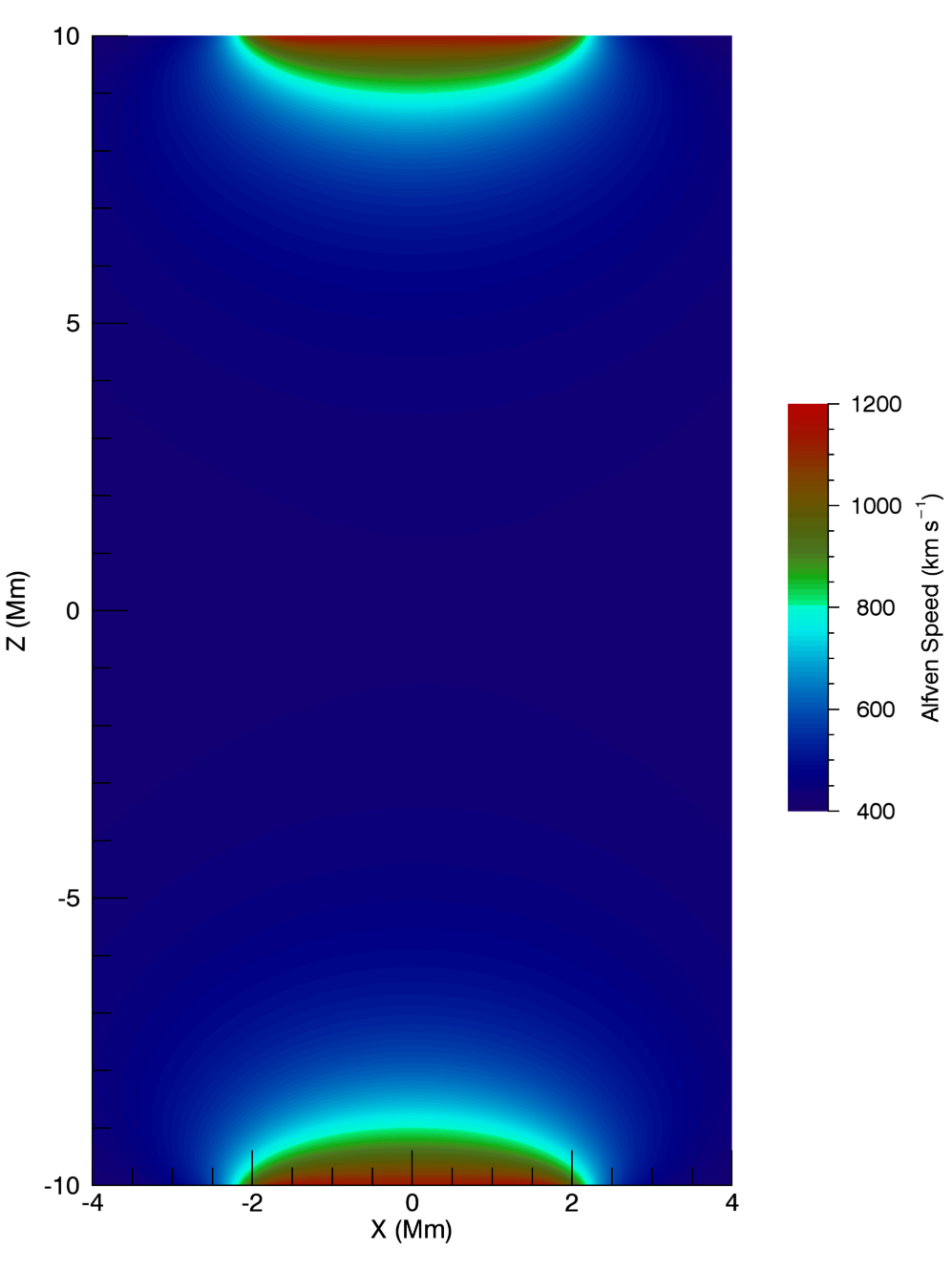}
  \caption{Alfv\'en speed in a vertical cut ($x=0$ plane) through the loop axis following the numerical relaxation.}
  \label{Alf_contour}
\end{figure}

The magnetic field lines (after relaxation) in the flux tube are displayed in Fig. \ref{Field_Profiles}. Here, we have traced field lines from $R=2$ Mm on the lower $z$ boundary of the computational box. We note that the displayed figure does not represent the entirety of the numerical domain and, as mentioned above, the flux tube expansion is not significantly restricted by boundary effects. Although the field expansion seems limited, we highlight that Fig. \ref{InField} (dashed line) shows that the field strength does decrease significantly at the loop apex. 

In Sect. \ref{Sec_Res}, we will often refer to the boundary of the expanded magnetic flux tube. However, we note that this is not well-defined following the numerical relaxation. Therefore, we will define the boundary of the flux tube to be the volume containing magnetic field lines that map to the (well-defined) boundary of the flux tube on the upper and lower $z$ boundaries of the domain.

In Fig. \ref{InField_Vertical} we consider the change in magnetic field strength along the loop axis that is caused by the expansion of the flux tube with height. Further, in Fig. \ref{Alf_contour}, we display a contour of the Alfv\'en speed in a cut along the longitudinal axis of the loop. We note that since the density is initially uniform, the Alfv\'en speed is simply proportional to the magnitude of the field. In both figures, we observe that the majority of the expansion occurs close to both foot points and the central portion (along the $z$ axis) of the loop is almost uniform. We highlight that at the loop apex, the field strength within the loop is close to that of the initial external field. Furthermore, the numerical relaxation conserves the azimuthal invariance of the flux tube.

\begin{figure}[h]
  \centering
  \includegraphics[width=0.5\textwidth]{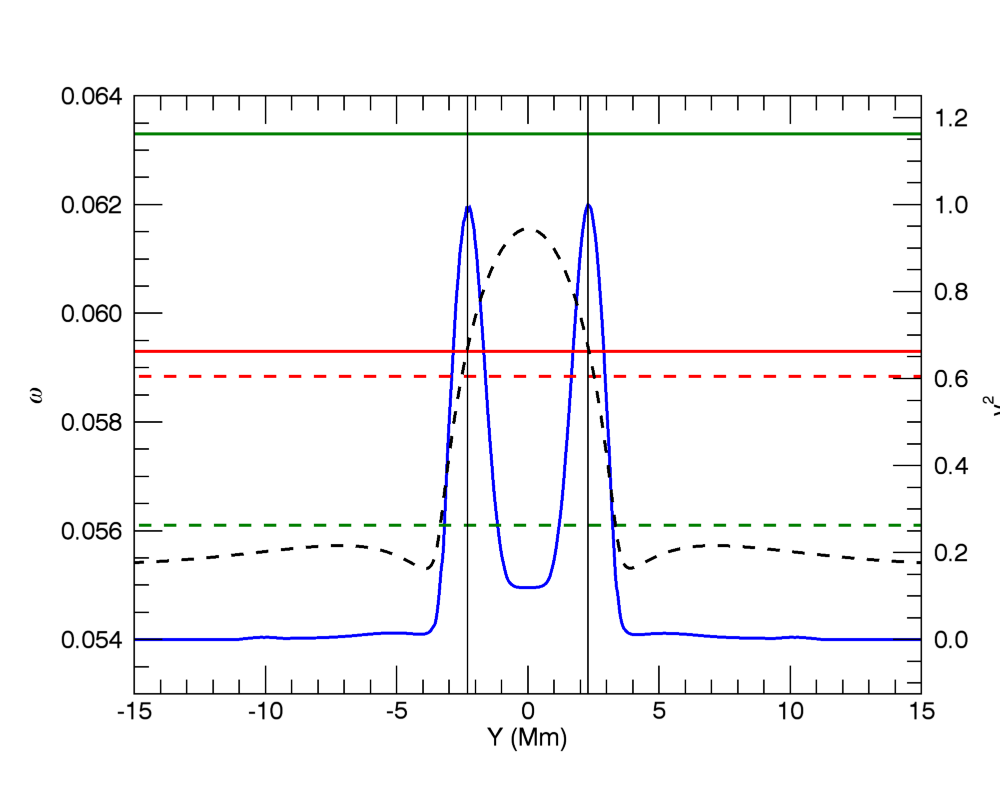}
  \caption{\emph{Dashed black line}: Natural Alfv\'en frequencies of field lines at the loop apex. \emph{Solid red line}: Kink frequency generated in the simulation. \emph{Green lines}: Theoretical estimates of the frequency of the fundamental kink mode using the kink speed, $v_k$, at the loop apex (dashed line) and tracking $v_k$ as a function of position along the loop axis (solid line). The estimate calculated using the kink speed at the loop foot points is significantly larger and is thus omitted from the figure. \emph{Dashed red line}: An additional estimate of the kink frequency calculated by solving an eigenvalue problem. \emph{Blue line}: The square of the azimuthal velocity, $v_{\phi}$, integrated in time over a wave period once significant energy has been transferred through resonant absorption ($t \approx 1500$ s). This is normalised to the maximum value observed. \emph{Solid black line}: The location of peak power for the azimuthal Alfv\'en wave.}
  \label{NatFreq}
\end{figure}

In Fig. \ref{NatFreq}, we show the natural Aflv\'en frequencies of field lines across the cross-section of the loop at the apex (dashed black line). In order to generate this plot, we solve equation 7 in \citet{Wright1994} for the frequency of the fundamental standing Alfv\'en wave along each field line in the $y=0$ plane. This equation accounts for the variation in Alfv\'en speed and the length of field lines. Additionally, the geometry of the flux tube is accounted for by a series of scale factors which may be interpreted physically in terms of the elemental separation of neighouring field lines. The scale factors are functions of position, and for an analytically defined equilibrium may have explicit expressions. Since our equilibrium is not analytical, we estimated the scale factors numerically by tracing magnetic field lines. The remaining curves in Fig. \ref{NatFreq} are discussed in the following section.

\begin{figure}[h]
  \centering
  \includegraphics[width=0.5\textwidth]{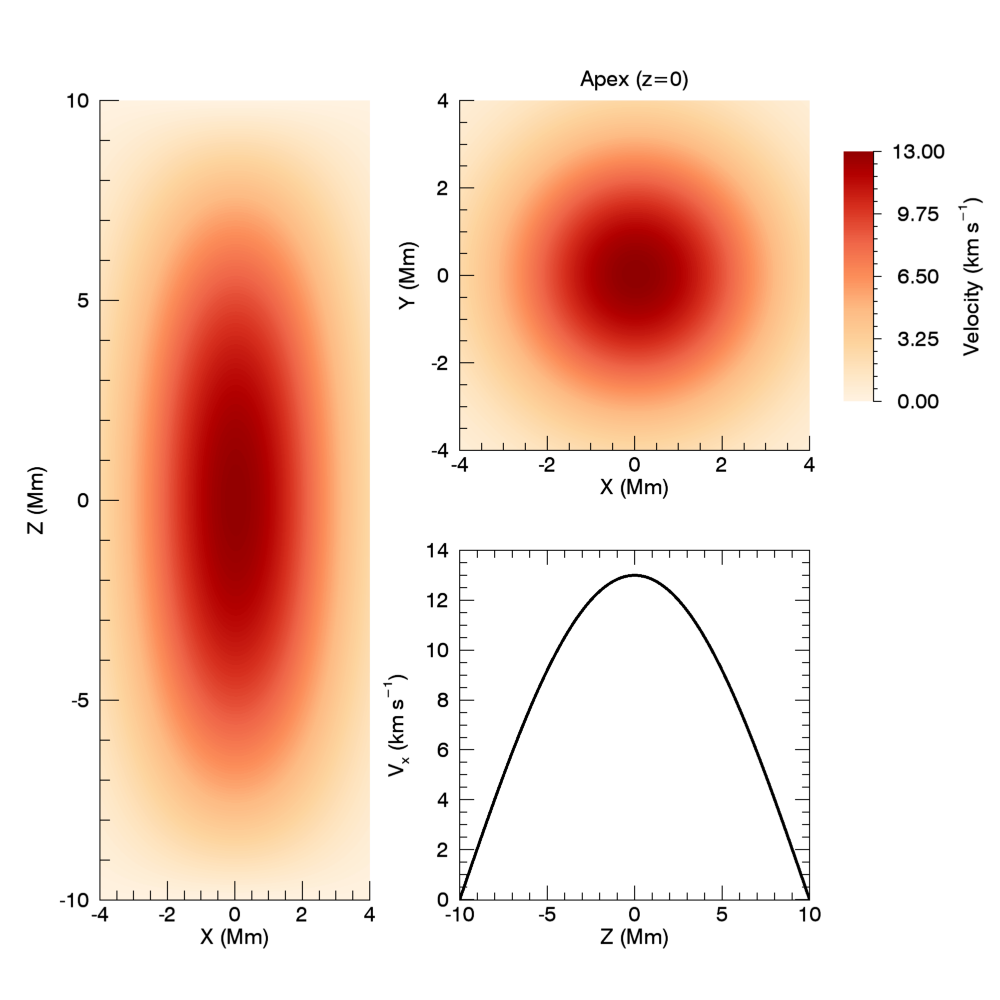}
  \caption{Initial velocity profile. Cuts in a vertical plane through the loop axis (left), a horizontal plane throught the loop apex (top right) and a plot of $v_x$ as a function of height along the loop axis (bottom right).}
  \label{InitVel}
\end{figure}

Following the numerical relaxation, we imposed a transverse velocity of the form $\vec{v}=\left(v_x, 0, 0\right)$ where

\begin{equation}
v_x = v_0e^{-\left(\frac{R}{r_w}\right)^2}\cos\left(\frac{\pi z}{2L}\right).
\end{equation}

Here, $v_0 \approx 13 \text{ km s}^{-1}$ is the maximum amplitude of the initial perturbation, $r_w$ is a parameter that ensures the width of the velocity profile is approximately the radius of the apex of the expanded flux tube (see Fig. \ref{Field_Profiles}) and $L = 20$ Mm is the length of the loop. The cosine term generates a fundamental standing mode with the velocity set to zero at the footpoints and maximal at the loop apex. 

In Fig. \ref{InitVel}, we show the initial velocity profile in vertical (left panel) and horizontal (upper right panel) cuts and along the central axis (lower right panel). Since the flux tube does not have a constant radius, this profile does not coincide with the width of the magnetic structure along the entire height of the flux tube. In particular, both the internal and external plasma is disturbed by this velocity perturbation. In this case, energy from both the internal and external plasma is readily transferred to Alfv\'en wave energy associated with resonant field lines. The main effect of the velocity profile not following the exact form of the magnetic flux tube is that additional harmonics (including radial wave modes) are excited. 

Despite this, throughout the simulations, we ensured that a node is located on both the upper and lower $z$ boundaries by enforcing zero velocities in these locations. Meanwhile, all other variables have zero gradients at the loop foot points. The $x$ and $y$ boundaries are periodic, however, in practice, flows are very small across these boundaries as a damping region is implemented at large $|x|$ and $|y|$ in order to minimise domain boundary effects on the oscillation. Throughout the duration of the simulation, the damping layers are well removed from the wave dynamics that are discussed hereafter. 

\section{Results}\label{Sec_Res}
Following the imposition of the initial velocity profile, a standing kink wave is generated. Magnetic tension, and to a much lesser extent magnetic (and gas) pressure gradients, act as the restoring forces. The observed period of the wave is approximately 106 s and thus, we find that the observed kink frequency is $\omega_k = 0.0593$. This corresponds to the solid red line in Fig. \ref{NatFreq}.

In a straight and slender flux tube with loop-aligned invariance, the kink speed, $v_k$ can be expressed as \citep[e.g.][]{Nakariakov2005}
\begin{equation}\label{kink_1} v_k = \sqrt{\frac{\rho_iv_{A,i}^2 + \rho_e v_{A,e}^2}{\rho_i + \rho_e}},
\end{equation}
where a subscript $i$ denotes a variable within the flux tube and a subscript $e$ denotes a variable within the external plasma. Although the current model violates the assumptions used to derive the above expression, we should be able to bound the expected period, $\tau_k$, using 
\begin{equation}
2 \int \dfrac{\mathrm{d} z}{\max{v_k}} \le \tau_k \le 2 \int \dfrac{\mathrm{d} z}{\min{v_k}}.
\end{equation}
Here, we have used a simple WKB approximation for the period of a standing wave mode with speed, $v_k$. The factor of 2 is included as the length of the loop is only half of the longitudinal wavelength for a fundamental mode. The integrals are calculated between the two nodes located at the upper and lower $z$ boundaries. Furthermore, we can estimate the expected period as 
\begin{equation} \label{kink_estimate}
\tau_k = \int \dfrac{\mathrm{d} z}{v_k(z)}.
\end{equation}
The three integrals in the preceeding equations provide an estimate and upper and lower bounds of the kink frequency. The upper bound is shown in Fig. \ref{NatFreq} as the solid green line. The dashed green line corresponds to the estimate calculated using equation \ref{kink_estimate}. The lower bound was found to be a frequency of 0.121 and for clarity is not included in Fig. \ref{NatFreq}. The solid red line corresponds to the actual frequency observed within the simulation ($\omega = 0.0593$).

We can provide an alternative estimate by solving the wave equation 
\begin{equation}
\dfrac{\partial^2 u}{\partial t^2} = v_k^2(z)\dfrac{\partial^2 u}{\partial z^2},
\end{equation}
{where $u$ is the perturbed component of the velocity, and by assuming a solution of the form
\begin{equation}
u = f(z) e^{i \omega t}.
\end{equation}
In order to solve the resulting eigenvalue problem, we shoot for the frequency that satisfies the boundary conditions and find an improved estimate of the kink frequency of 0.0588. This is displayed as the dashed red line in Fig. \ref{NatFreq}.

\begin{figure}[h]
  \centering
  \includegraphics[width=0.5\textwidth]{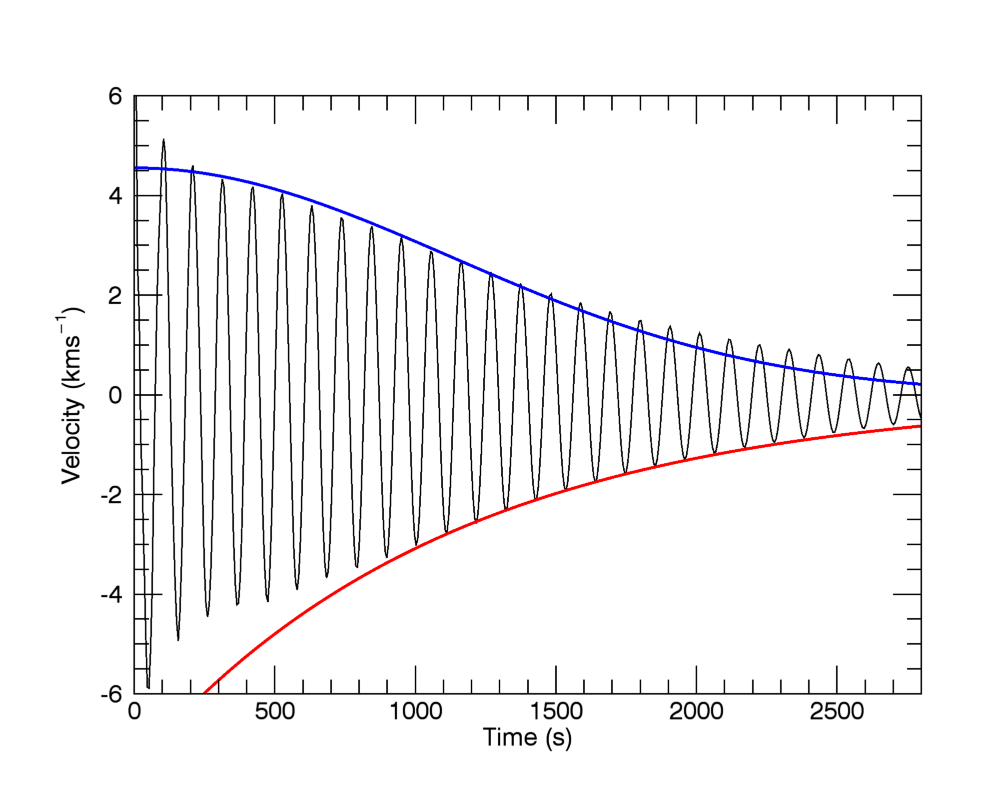}
  \caption{Velocity at the centre of the flux tube cross-section indicating the damping rate of the kink mode. Two damping profiles are also displayed. Upper: Gaussian (blue) profile. Lower: Exponential (red) profile.}
  \label{Damping}
\end{figure}

\begin{figure}[h]
  \centering
  \includegraphics[width=0.32\textwidth]{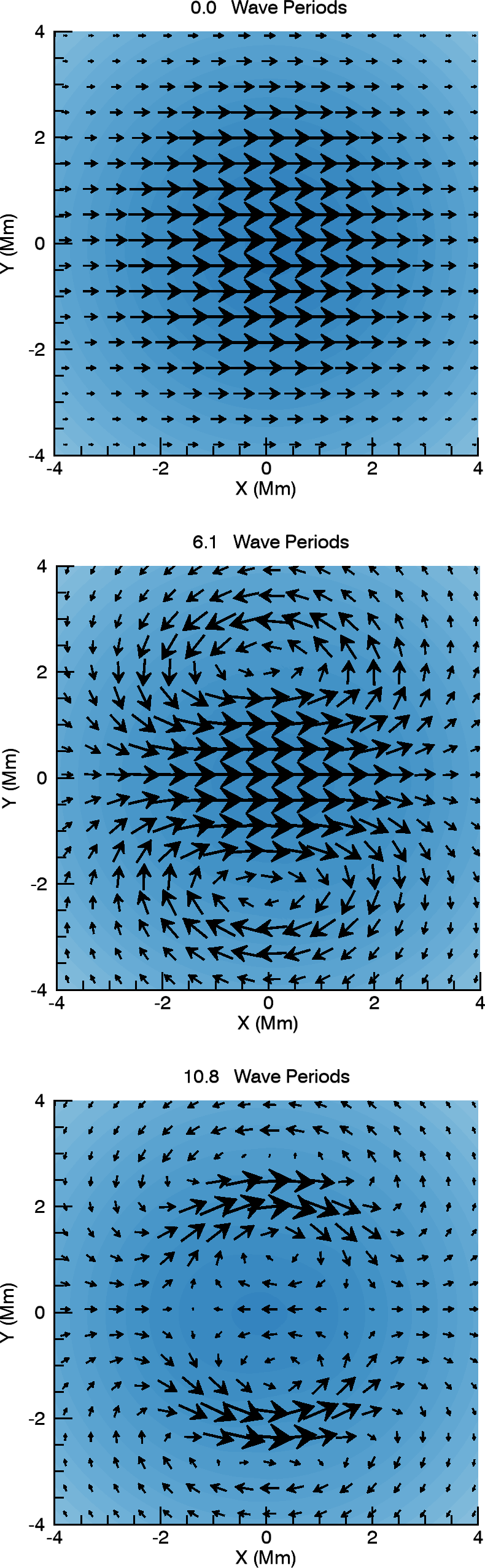}
  \caption{Velocity vectors in the horizontal plane at the loop apex. Three times are shown: upper - initial velocity profile, middle - dipole flow has formed, lower - kink mode energy has been transferred into azimuthal Alfv\'en wave.}
  \label{VelVec}
\end{figure}
 
\subsection{Resonant Absorption}
Even in ideal conditions, the kink wave experiences damping at a significantly higher rate than can be accounted for by the small amount of (numerical) dissipation inherent to the simulation. In Fig. \ref{Damping}, we plot the velocity in the centre of the flux tube at the loop apex ($x=y=z=0$) as a function of time. We note that the initial amplitude of $13 \text{ km s}^{-1}$ decays very quickly (within 1 wave period) to approximately $5 \text{ km s}^{-1}$. For clarity, this is not shown in Fig. \ref{Damping} and is associated with a short-lived leaky mode \citep[e.g.][]{Terradas2006b} that transfers energy into the surrounding plasma until the kink mode is established.

We observe the subsequent damping of the kink mode over many wave periods and have included two decay profiles; the more typical exponential decay (red) and a Gaussian curve (blue). We note that the Gaussian fit is more suitable until around $t=1200$ s, at which time the exponential becomes more appropriate. This is in agreement with \citet{Pascoe2013} in which the authors demonstrate that resonant absorption is associated with an initial phase of Gaussian damping before exponential damping dominates at a later time. 

As discussed above, the process of resonant absorption transfers energy from the kink mode into energy associated with azimuthal Alfv\'en waves. This is readily observed in the velocity field displayed in Fig. \ref{VelVec}. In the first panel we show the initial velocity profile (corresponds to Fig. \ref{InitVel}). This demonstrates that the majority of the initial kinetic energy is contained within the central region of the flux tube. As this plasma moves, it is replaced by external plasma and thus a dipole flow forms immediately. This persists over many wave periods (second panel) as the kink mode decays. Throughout the simulation, resonant field lines are excited by the kink mode and begin to oscillate as part of an azimuthally polarised Alfv\'en wave which can be observed in the third panel. 

\begin{figure}[h]
  \centering
  \includegraphics[width=0.5\textwidth]{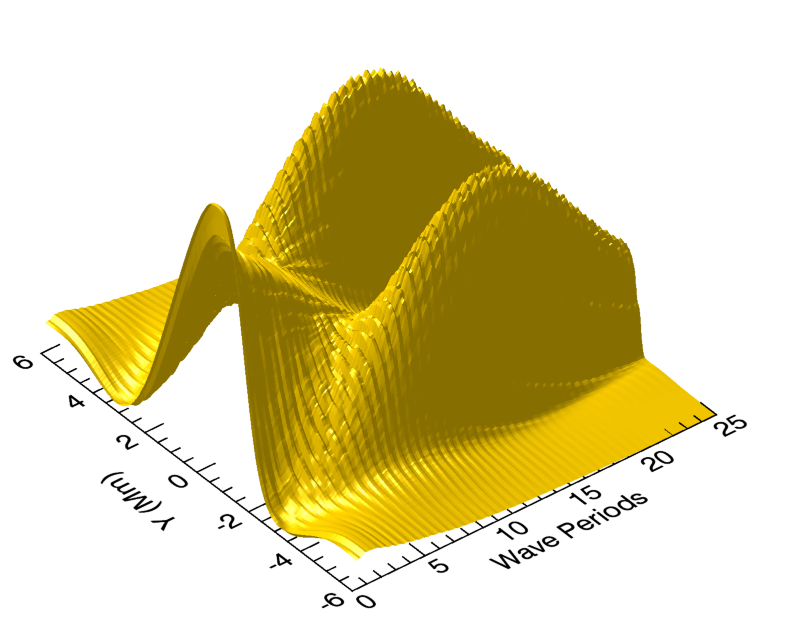}
  \caption{Azimuthal kinetic energy in the line $x=z=0$ as a function of time.}
  \label{WavePower}
\end{figure}

\begin{figure*}[h]
  \centering
  \includegraphics[width=0.87\textwidth]{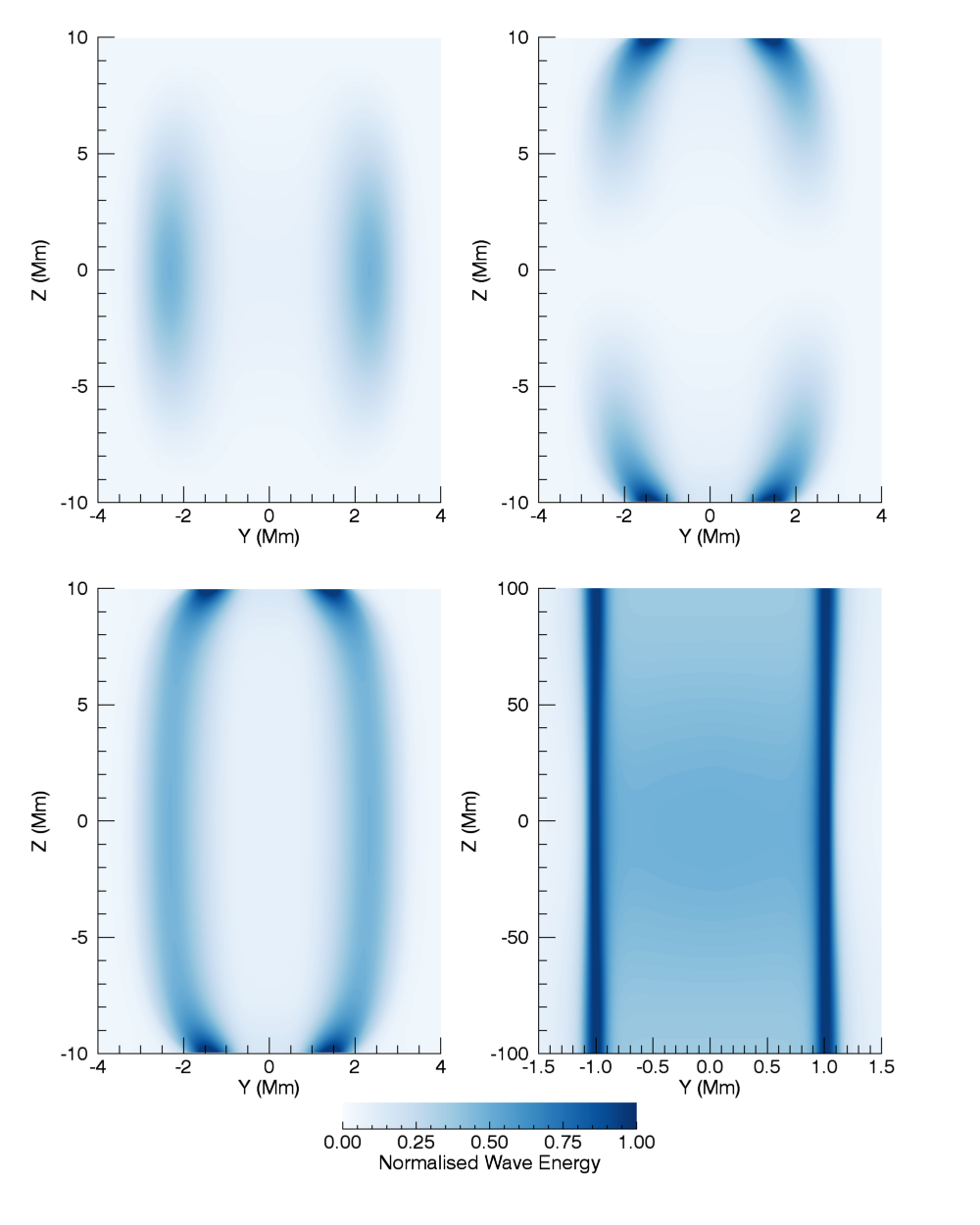}
  \caption{Vertical cuts through the loop axis showing the vertical structure of the Alfv\'en wave. The upper two panels show the kinetic (left) and magnetic (right) wave energy, integrated in time over a full period. The lower left panel is the sum of the two components of wave energy and the lower right panel is an analogous plot using data from straight field (no magnetic field expansion) experiments. In each case, we have normalised using the total Alfv\'en wave energy present in the respective simulation.}
  \label{AlfStruct}
\end{figure*}

This mode conversion transfers wave energy from the central region of the flux tube, to the boundary between the internal and external plasma. This boundary is well-defined close to the loop foot points, however, due to the expansion of the flux tube, it is much less apparent close to the loop apex. Indeed, comparing the solid and dashed line in Fig. \ref{InField}, demonstrates the change of the boundary region between the loop foot points and the loop apex. We highlight the result that resonant absorption can still occur even if there is very little gradient in Alfv\'en speed along large sections of a magnetic flux tube.

In Fig. \ref{WavePower}, we show how the location of kinetic energy along the line $x=z=0$ (a vertical line when viewed in the panels in Fig. \ref{VelVec}) changes throughout the course of the simulation. Initially, it is predominantly located within the core of the loop (kink wave), however, as time progresses, it is transferred to two narrow layers within the boundary region of the flux tube (Alfv\'en wave). The $y$ locations of the peak power (for both wings of the Alfv\'en wave) beyond a time of approximately 6 wave periods, indicate the radial position of the resonant field lines. The azimuthal wave power then begins to decrease once the majority of the kink mode energy has been exhausted and as the weak viscosity (user-imposed and, to a lesser extent, due to numerical effects) dissipate the wave energy.   

In Fig. \ref{NatFreq}, we include the location of Alfv\'en wave power (blue line) as a function of distance from the centre of the loop. We show the magnitude of $v^2$ (this is approximately proportional to the kinetic energy as the density is largely constant) integrated in time over the duration of a wave period. We select a time once significant mode conversion has occurred ($t \approx 1500$ s). At this time, it is clear that the wave power peaks in the boundary of the magnetic structure. In Fig. \ref{NatFreq}, the vertical black lines indicate where the kink frequency observed within the simulation matches the theoretically predicted Alfven frequency. The fact that this location coincides with the peaks of $v^2$ confirms our interpretation in terms of resonantly excited Alfven waves driven by the kink mode.

\subsection{Alfv\'en Wave Structure}
On account of the expansion of the flux tube with height, a narrow resonant layer of field lines at the loop foot point, maps to a much wider layer close to the loop apex. Accordingly, the width of the resonant layer varies with position along the flux tube. In particular, it exists over a much smaller horizontal extent close to the upper and lower $z$ boundaries than at the loop apex.

In Fig. \ref{AlfStruct}, we display the spatial profile of Alfv\'en wave power once significant energy has been transferred from the kink wave ($t \approx 1500$ s). The upper panels correspond to the kinetic (left) and magnetic (right) wave energy, integrated over a wave period. The lower left panel is simply a sum of the two upper panels. The lower right panel, on the other hand, corresponds to the simulations presented in \citet{Howson2017a} in which a straight (no magnetic field expansion) flux tube is considered and the resonant absorption is associated with a non-uniform density profile. As with the lower left panel, it shows the sum of the magnetic and kinetic wave energy integrated over a wave period. We note that the flux tube is significantly longer than in the case presented within this article. For comparison, in each panel, we have normalised the total Alfv\'en wave energy in the corresponding simulation. 

In the top left panel of Fig. \ref{AlfStruct}, we observe that the magnitude of the kinetic energy is largest at the loop apex which coincides with the location of the velocity antinode in the fundamental mode. Meanwhile, the magnetic component (upper right panel) of the wave energy is largest at the loop foot points as this is where the perturbation of the magnetic field is greatest. Since we generate a standing wave, the velocity and magnetic field perturbations are out of phase, however, integrating over a wave period means we might expect the wave energy to be approximately constant along a field line. 

This is indeed the case in the straight flux tube case (lower right panel) as the width of the resonant layer does not change along the length of the loop axis. At a radius of $R \approx 1$ Mm, we see that the time-integrated Alfv\'en wave energy is approximately constant with height. However, this is not the case in the boundary of the flux tube in the expanding field simulation, even when the shape of the field lines is accounted for. The density of field lines increases near the loop foot points and this has the effect of concentrating wave energy close to the upper and lower $z$ boundaries (see lower left panel).

At this stage, we note that the straight field cases can become unstable to the Kelvin-Helmholtz instability (KHI). This is due to the radial shear in the velocity that is associated with the excited azimuthal Alfv\'en waves. The lower right panel of Fig. \ref{AlfStruct} corresponds to a time prior to the formation of the instability. The development of the KHI leads to a deformation and expansion of the loop's cross-section, particularly at the loop apex \citep[e.g.][]{Terradas2008a, Antolin2014, Magyar2015}. This expansion will increase the width of the flux tube boundary and thus, can induce similar effects to the expanding field case. The KHI does not form in the expanding field case as the radial velocity shear that develops is much smaller than in the straight field simulation. This is because the width of the resonant layer is much larger and because resonant absorption proceeds at a reduced rate in this case (see Fig. \ref{Herring_bone} and associated discussion).

Whilst the straight field profile is able to sustain a narrow shell of resonant field lines along the entire length of the flux tube (prior to the development of the KHI), the field expansion considered here results in the resonant layer being much wider in comparison to the loop length. The width of this resonant layer as a function of height is shown in Fig. \ref{AlfWidth} for the expanding field case. Here, the width is calculated using the full width at half maximum of the azimuthal velocity profile for one wing of the Alfv\'en wave at each height within the domain. We observe that since most of the expansion of the flux tube occurs close to the foot points (see Figs. \ref{InField_Vertical} \& \ref{Alf_contour}), the greatest change in width of the resonant layer occurs closest to the upper and lower boundaries of the domain.

\begin{figure}[h]
  \centering
  \includegraphics[width=0.5\textwidth]{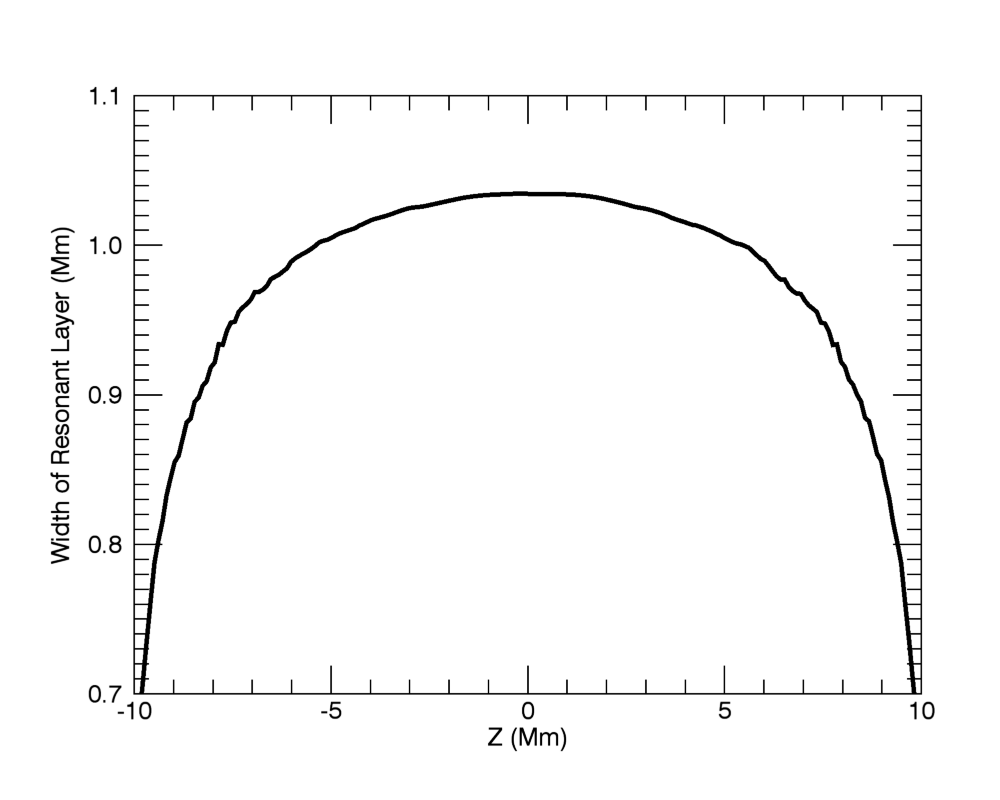}
  \caption{Width of the Alfv\'en wave as a function of height along the flux tube.}
  \label{AlfWidth}
\end{figure}

The smaller (in relation to the kink mode) length scales of the Alfv\'en wave are important in the context of wave energy dissipation, and hence coronal heating. In both simulations, the gradients in the velocity and magnetic fields are larger for the localised wave mode than for the global (kink) mode. As such, the effects of viscosity (on the velocity field) and resistivity (on the magnetic field) are more significant for the Alfv\'en wave than for the initial kink mode. However, since the set-up discussed within this article is associated with a much wider resonant layer (due to the field expansion), the gradients in the magnetic and velocity fields are smaller than in the straight field simulation.

In general, due to the field expansion, the length scales remain larger in the current model than in the classical straight field case. As a result, in a comparable straight field simulation, we expect more efficient wave heating due to phase mixing and the development of dynamic instabilities such as the Kelvin-Helmholtz instability \citep[e.g.][]{Terradas2008a, Karampelas2018}. However, a rigorous consideration of the system in a non-ideal regime is beyond the scope of this publication and will instead be considered in subsequent work.

\begin{figure}[h]
 \centering
  \begin{subfigure}{0.45\textwidth}
  \includegraphics[width=\linewidth]{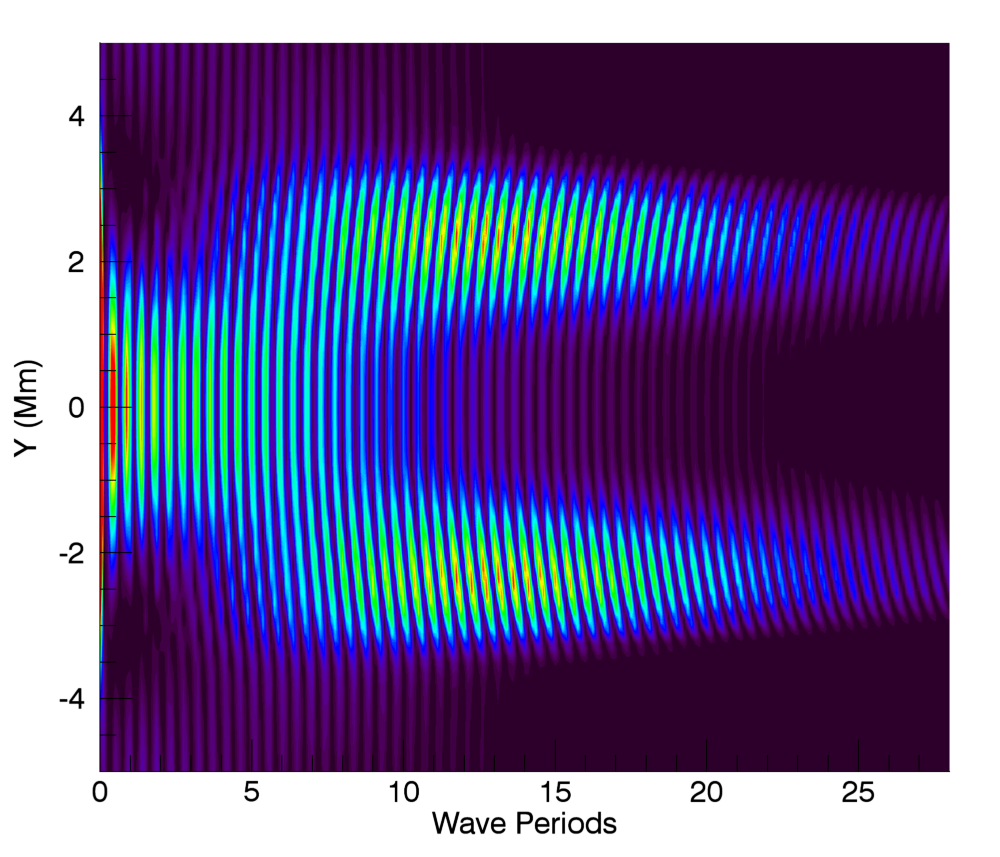}
  \caption{Expanding magnetic field.} \label{Herring_bone_a}
\end{subfigure}
\begin{subfigure}{0.45\textwidth}
  \centering
  \includegraphics[width=\linewidth]{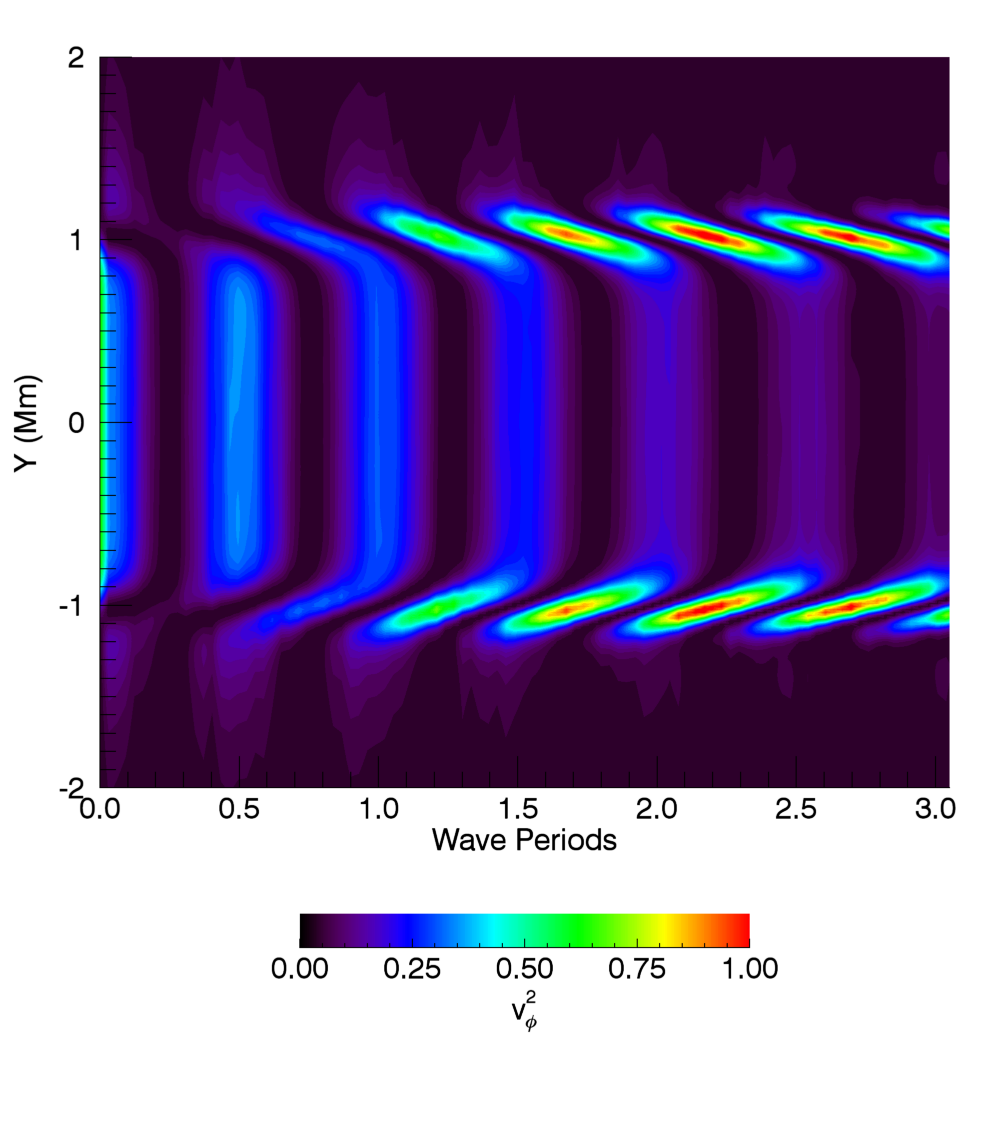}
  \caption{Uniform magnetic field.}
  \end{subfigure}%
  \caption{Transfer of wave energy during resonant absorption. We show the square of the azimuthal velocity, $v_{\phi}^2$, along a diameter through the loop apex and as a function of time. In each case, we have normalised by the maximum value observed during the simulation.}
  \label{Herring_bone}
\end{figure}

\begin{figure*}
\centering
\begin{subfigure}[b]{0.48\textwidth}
\centering
   \includegraphics[width=1\linewidth]{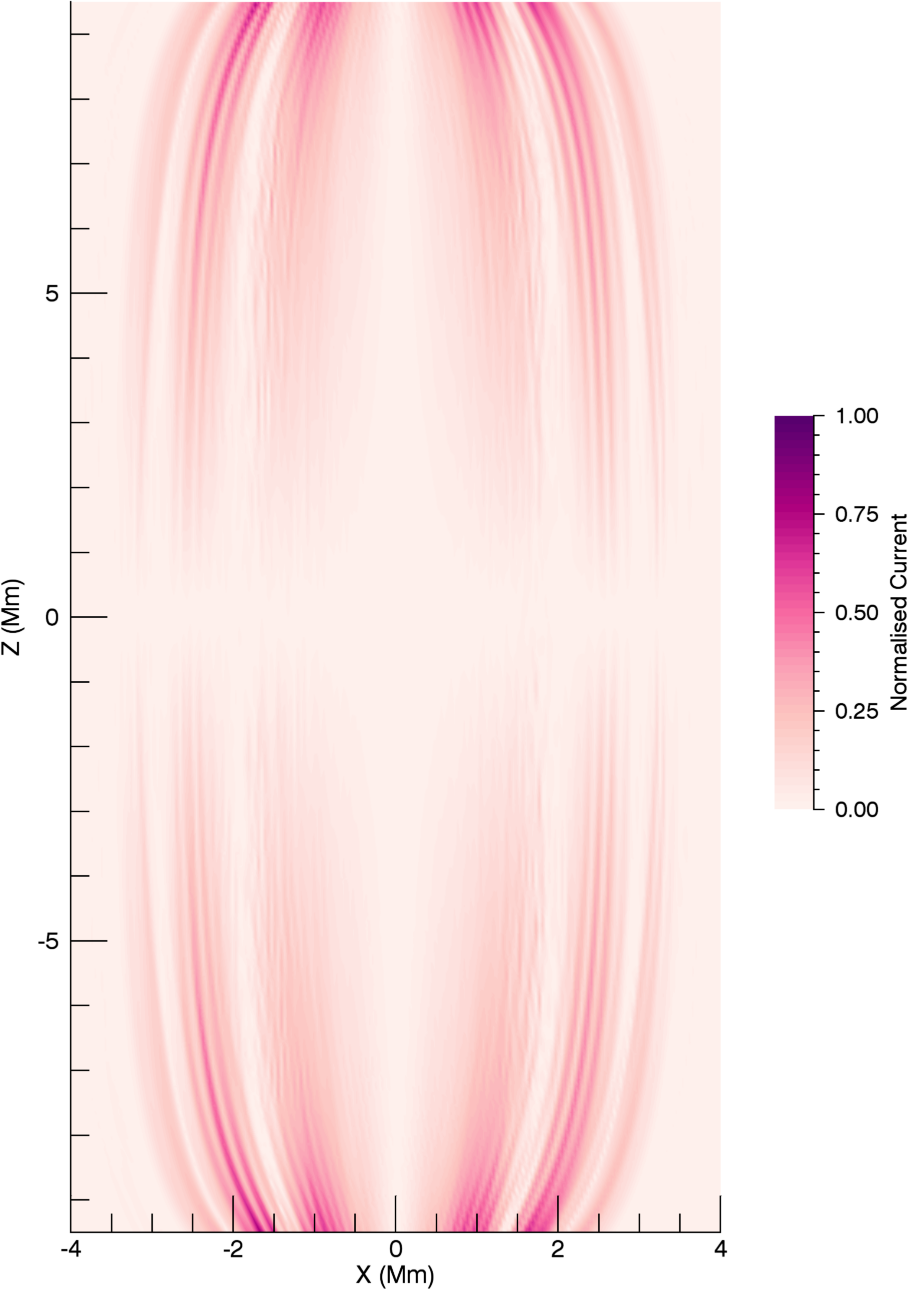}
     \caption{Current.}
   \label{AlfCur}
\end{subfigure}
\begin{subfigure}[b]{0.48\textwidth}
\centering
   \includegraphics[width=1\linewidth]{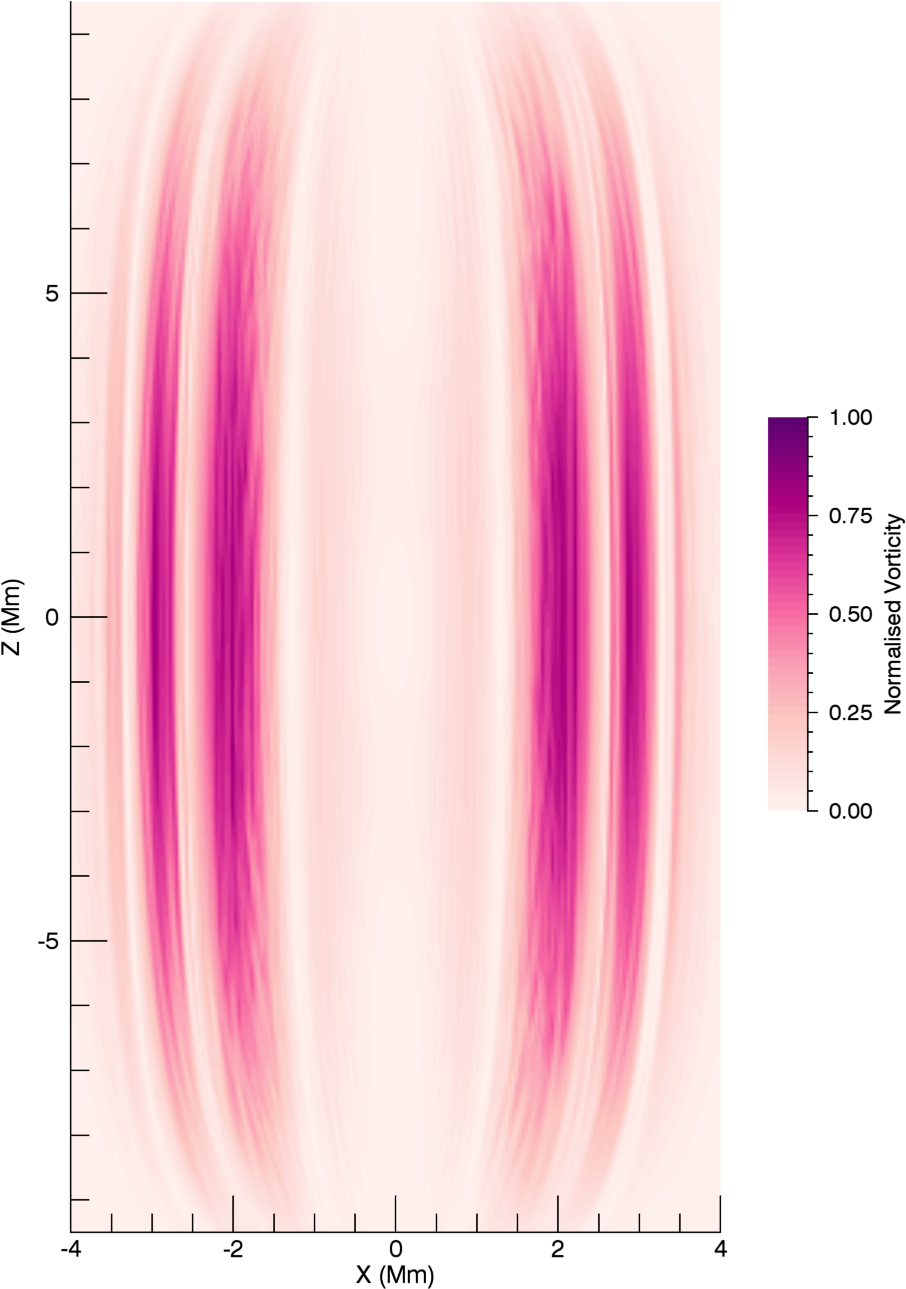}
  \caption{Vorticity.}
  \label{AlfVor}
\end{subfigure}
\caption{Magnitude of the current density, $|\vec{j}|$, and vorticity, $|\vec{\omega}|$ associated with the azimuthal Alfv\'en wave in the $y=0$ plane at $t \approx 1500$ s.}
  \label{AlfCurAndVort}
\end{figure*}

In Fig. \ref{Herring_bone}, we show how the location of kinetic energy along the line $x=z=0$ changes throughout the course of the simulation. We show the results for the expanding field simulation described within this article (upper panel) and, for comparison, the straight field simulation outlined in \citet{Howson2017a}. The upper panel can be easily compared to Fig. \ref{WavePower}. Since the observed kink frequency is different in both experiments, we display the energy transfer as a function of the number of wave periods.

In both simulations, we see that energy is transferred from the core region to the boundary of the loop as time progresses. It is clear that resonant absorption occurs much more quickly (in terms of wave periods) in the straight field simulation. This is intuitive given that the radial Alfv\'en speed gradient (a critical requirement for resonant absorption to proceed) is significant along the entire length of the flux tube in the straight field case, but not in the expanding field structure. Indeed, in the simulation described within this article, the Alfv\'en speed is almost constant throughout the cross-section at the loop apex (see Fig. \ref{Alf_contour}). Importantly, there is a larger contrast between the natural Alfv\'en frequencies of field lines inside and outside of the flux tube in the straight field case than in the expanding magnetic field simulation.

Another important difference between the two simulations is highlighted by considering the relative internal and external Alfv\'en frequencies. In the density-defined loops, the internal frequency is lower than the external frequency. However, this is reversed in the case of flux tubes defined by a magnetic field enhancement. This can be observed by considering the orientation (direction) of the wave fronts in the two panels of Fig. \ref{Herring_bone}. In the upper panel, an Alfv\'en wave front appears to propagate away from the centre of the loop, whereas in the lower panel, this behaviour is reversed. In the flux tube with the density enhancement (lower panel), the Alfv\'en speed (and thus the natural frequency) within the loop is lower than in the exterior plasma. Hence, a standing Alfv\'en wave will be first observed on field lines at the edge of the boundary region. The lower frequency field lines closer to the loop centre will oscillate at a slightly later time, and hence the wave front appears to propagate towards the centre of the loop. In the simulation corresponding to the upper panel, the natural Alfv\'en frequencies of field lines within the flux tube are higher than in the external plasma and thus, the opposite effect is observed. This inversion (with respect to the typically modelled case) might be expected in chromospheric flux tubes (if the internal frequency is higher than the external frequency) but is unusual for a simulation concerning coronal structures.

\subsection{Current and Vorticity}
The small scales associated with the Alfv\'en wave manifest in the form of currents (for the magnetic field) and vorticities (for the velocity field). The radial non-uniformity in natural Alfv\'en frequency will induce out-of-phase wave behaviour on neighbouring field lines, leading to an enhanced rate of dissipation through phase mixing \citet{Heyvaerts1983}. 

In Fig. \ref{AlfCurAndVort}, we display the currents (left-hand panel) and vorticities (right-hand panel) associated with the azimuthal Alfv\'en waves. To generate these plots, we consider a time after a significant amount of energy has been transferred from the global wave to the localised modes ($t \approx 1500$ s). The shape of the expanded flux tube remains apparent in these plots and corresponds to the Alfv\'en wave structure (see Fig. \ref{AlfStruct}).

\begin{figure}[h]
  \centering
  \includegraphics[width=0.5\textwidth]{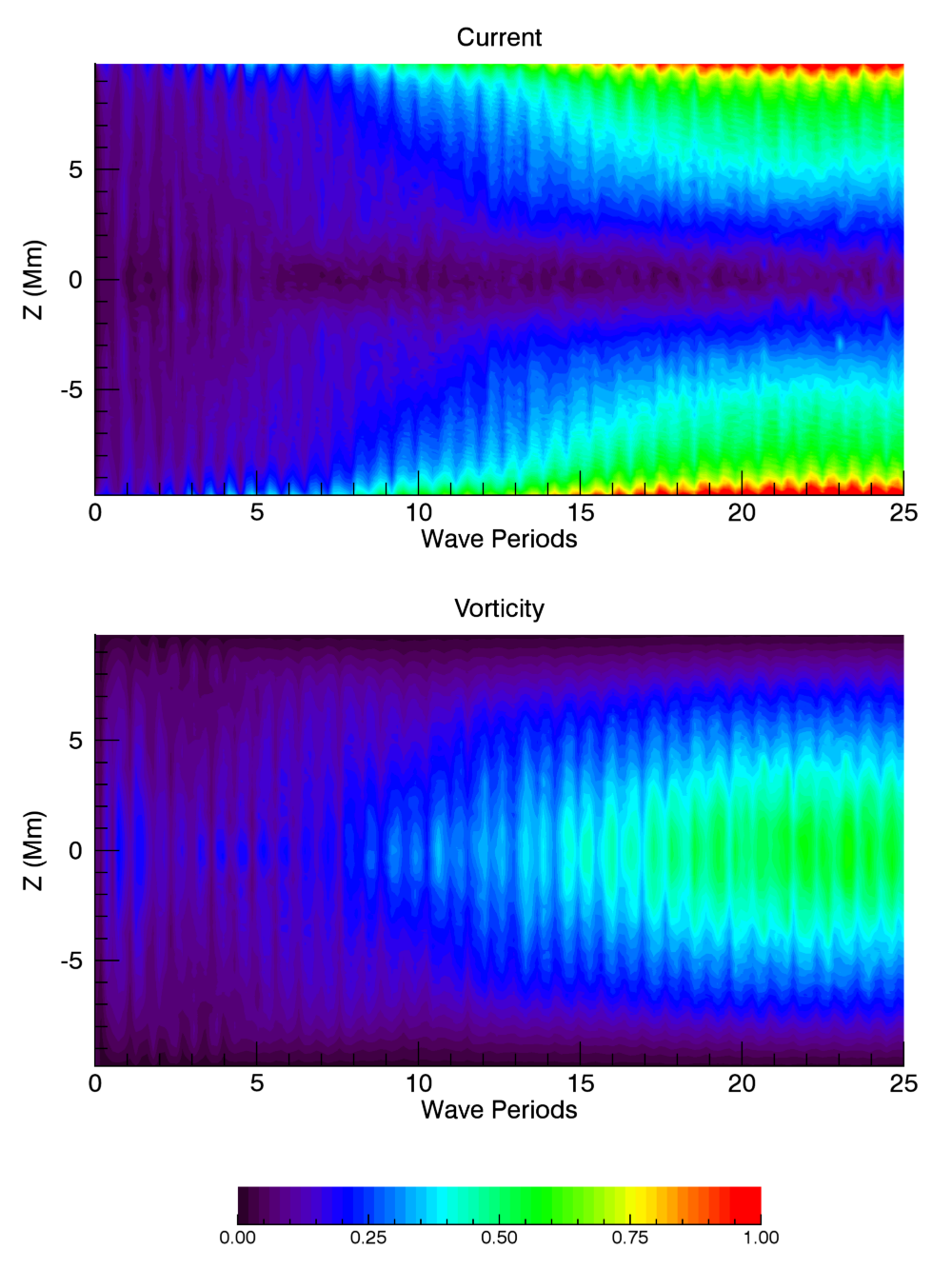}
  \caption{Total current (upper) and vorticity (lower) across loop cross-section as a function of distance along the loop and time. In both cases, we have normalised by the maximum value of the total current.}
  \label{CurVortGrow}
\end{figure}

As discussed previously, over the course of a wave period, the energy in a standing Alfv\'en wave is partitioned into kinetic energy (located close to the antinode) and magnetic energy (located close to the nodes). This ensures that the largest currents form close to the foot points of the magnetic flux tube and the largest vorticities form close to the loop apex. This phenomenon is independent of the form of the flux tube and the significance for the spatial distribution of wave heating is discussed in \citet{Karampelas2017}. In this case, it can be observed by comparing the location of the largest currents and vorticities in Figs. \ref{AlfCur} \& \ref{AlfVor}, respectively. 

The large-scale, strand-like structures that appear in both contour plots are indicative of phase mixing. The radial non-uniformity in Alfv\'en frequency (see Fig. \ref{NatFreq}) ensures that Alfv\'en waves on neighbouring radial shells gradually become out of phase. This, in turn, generates small scales across the resonant layer and, in a non-ideal regime, wave energy would be dissipated across this region of the flux tube.

In addition to the longitudinal out-of-phase behaviour between the currents and vorticities (described above), in Fig. \ref{AlfCurAndVort}, we observe that the phase-mixing strands are also out-of-phase radially. This is simply due to the magnetic and kinetic energy being out-of-phase in a standing Alfv\'en wave. On field lines with large currents, most of the wave energy is magnetic in nature and, on the other hand, for field lines with large vorticities, most of the wave energy is kinetic.

As resonant absorption transfers energy from the global mode to localised waves, the magnitude of currents and vorticities within the numerical simulation will increase. This can be observed in Fig. \ref{CurVortGrow} for $\vec{j}$ (upper panel) and $\vec{\omega}$ (lower panel), respectively. In each case we integrate the magnitude of the vectors over the loop cross-section at each height and display this quantity as a function of time. We have normalised both quantities by the maximum of the total current and thus we see that the current attains larger values than the viscosity.

In both cases, the formation of small scales as resonant absorption progresses can be observed and once again we note that vorticity dominates at the loop apex and currents dominate at the loop foot points. The small peak in vorticity at the beginning of the simulation corresponds to velocity gradients associated with the initial kink mode. Since the magnetic energy is concentrated over a smaller region than the kinetic energy (see Fig. \ref{AlfStruct}), we anticipate that for comparable values of resistivity and viscosity, Ohmic heating will be the more significant dissipation mechanism. 

\section{Discussion and Conclusions} \label{Discussion}
Within this paper, we have presented a model of an expanded magnetic flux tube excited with a standing, transverse, kink oscillation. In agreement with existing models, the presence of a non-uniform, transverse profile in the Alfv\'en frequency, permits resonant absorption to augment the decay of the fundamental kink mode as energy is transferred to localised, azimuthal Alfv\'en waves. 

Previously, many models have considered the effects of a density enhancement within the coronal loop. However, we have presented a departure from the typical model in the sense that the flux tube we have studied is defined by a magnetic field strength enhancement and the density is initially uniform throughout the numerical domain. In this instance, the transverse profile in the natural frequency is associated with both the difference between internal and external magnetic field strengths and the variation in the length of field lines. Additionally, in contrast to many previous studies, the ratio between the internal and external Alfv\'en speeds is not constant along the length of the flux tube. Consequently, the applicability of previous analytic studies to this model is limited.


As with straight flux tube simulations, the azimuthal waves that are excited during resonant absorption exist over much smaller spatial scales than the global mode and are thus associated with larger gradients in the magnetic and velocity fields. These gradients correspond to currents and vorticities that increase in magnitude as the resonant absorption and phase mixing progresses. However, since the resonant layer is much wider than in typical straight flux tube studies (prior to the formation of the Kelvin-Helmholtz instability), the transverse length scale of the Alfv\'en wave remains much larger in the current model. Since we see that both resonant absorption and phase mixing progress at slower rates in the expanding field model, in a non-ideal regime, we expect wave heating to be less efficient in this case. Due to the nature of the standing Alfv\'en wave, the largest currents form at the loop foot points and the largest vorticities at the loop apex. The expansion of the flux tube ensures that the width of the resonant layer is much smaller close to the foot points than at the loop apex. Hence, given comparable resistivity and vorticity coefficients, in a non-ideal regime we can expect Ohmic heating to be the dominant cause of wave energy dissipation \citep{VanDoors2007} for this model.

In this article, we have studied a standing kink mode in a relatively short (for the solar atmosphere) magnetic flux tube. As such, it may be more representative of transition region loops which have been observed by \citet{Hansteen2014}, for example. Indeed, this model may be particularly applicable to short loops which exhibit large expansion at low altitudes. Since the radial Alfv\'en frequency gradient, depends on the amount of flux tube expansion, we expect that longer, coronal loops will exhibit a slower rate of resonant absorption and phase mixing than observed in this model. 

The absence of any density enhancement implies that the waves described within this publication would be very difficult to identify even with the increased detection power provided by contemporary observational instruments. Indeed, the initial flux tube is invisible to all but sensitive magnetic field measurements which are not currently possible within the coronal volume. Despite this, the next generation of solar telescopes such as DKIST, will hopefully provide insight into the nature of such magnetic structures within the Sun's atmosphere. Detecting the wave itself may be possible using Doppler velocities, however, favourable conditions are required as flows within dense structures along the line of sight will likely dominate any observed signal.

Despite these observational difficulties, it may be expected that such flux tube structures exist throughout the corona. Magnetic field within the outer solar atmosphere is typically connected to small scale flux patches in the photosphere/chromosphere and as the field enters a low plasma-$\beta$ regime, in order to maintain an equilibrium, it must become approximately force free. Thus, if we assume low levels of magnetic twist, it is reasonable to expect the field to expand significantly with height. As we have shown, in order to fully explain the damping behaviour of fundamental standing waves at high altitude, the expansion of the magnetic field closer to the solar surface should be considered. In particular, the global frequency of the field line is critically important and cannot simply be inferred from the local frequency at the loop apex. It is not possible to accurately predict the decay of a standing kink mode unless the internal and external Alfv\'en speeds are well constrained along the entire length of the flux tube. 

A major criticism of previous wave heating models \citep[see e.g.][]{Cargill2016}, is that the density profile typically assumed for resonant absorption/mode coupling and phase mixing models cannot be generated, or sustained, by the dissipation of MHD waves alone. Despite this, other authors have argued that wave heating models will dissipate energy across the cross-section of a coronal loop \citep[e.g.][]{Ofman1998, Karampelas2018}. The model presented within this paper provides a proof of principle suggesting that the density profile is not essential for wave heating to occur. Furthermore, any substantial wave heating in such a flux tube could lead to the evaporation of chromospheric plasma into the corona and the generation of a new density structure which would complicate the wave dynamics and may be observed as a new coronal loop.

\vspace{1cm}

{\emph{Acknowledgements.}} The research leading to these results has received
funding from the UK Science and Technology Facilities Council (consolidated grant ST/N000609/1), the European Union Horizon 2020 research and innovation programme (grant agreement No. 647214). IDM received funding from the Research Council of Norway through its Centres of Excellence scheme, project number 262622. PA acknowledges funding from his STFC Ernest Rutherford Fellowship (grant agreement No. ST/R004285/1). TVD was supported by funding from the European Research Council (ERC) under the European Union’s Horizon 2020 research and innovation programme (grant agreement No. 724326). ANW was partially supported by the Leverhulme Trust (through research grant RPG-2016-071). 

This work used the DiRAC Data Analytic system at the University of Cambridge, operated by the University of Cambridge High Performance Computing Service on behalf of the STFC DiRAC HPC Facility (www.dirac.ac.uk). This equipment was funded by BIS National E-infrastructure capital grant (ST/K001590/1), STFC capital grants ST/H008861/1 and ST/H00887X/1, and STFC DiRAC Operations grant ST/K00333X/1. DiRAC is part of the National e-Infrastructure.

\bibliographystyle{aa}        
\bibliography{MMC.bib}           

\end{document}